\documentstyle[prl,aps,epsf]{revtex}
\begin{document}
\draft

\narrowtext
\twocolumn[\hsize\textwidth\columnwidth\hsize\csname@twocolumnfalse%
\endcsname


]
{\bf Comment on ``Impurity states and the absence of
quasiparticle localization \\ in disordered $d-$wave
superconductors''}

In a recent Letter \cite{BS} Balatsky and Salkola made
their conclusion about the absence of localization in disordered 
$d-$wave superconductors exploiting the idea 
that strongly overlapping impurity states formed a network
capable to provide the transport. 
We present here the correct estimate of hopping matrix  
elements $\widetilde V_{ij}$ between two impurities which were the 
basic quantity for the subsequent consideration in \cite{BS}. 
These elements were given by $\widetilde V_{ij}\propto\widehat 
G^{0}({\bf r})$ with the electron Green function $ \widehat 
G^{0}({\bf r}) = \int d^2k e^{i{\bf kr}} \xi_k/(\xi_k^2+|
\Delta(\varphi)|^2)$, where $\xi_k = k^2/2m - \varepsilon_F$ and 
$\Delta(\varphi) = \Delta_0 \cos2\varphi$ is the $d-$wave 
superconducting order parameter. Performing the $\xi_k$-integration 
we get
\begin{equation}  
\widetilde V_{ij}\propto\frac{k_F}{v_F}{\rm Im}
\int_{-\pi/2}^{\pi/2}
d\varphi e^{ik_{F}r\cos\varphi} 
e^{-\frac{r}{v_{F}}\cos\varphi|\Delta(\varphi-\varphi_r)|}.
\end{equation}
Assuming that $|\varphi|\lesssim 1$ and $\varphi_r$ is measured 
from the node and small, one can expand the exponents in (1). 
After rescaling the variable 
$\varphi\to\phi=\varphi\sqrt{\frac{1}{2}k_F r}$ it yields
\begin{equation}\displaystyle{
\widetilde V_{ij} \propto \frac{1}{\sqrt{k_F r}} {\rm Im} 
\left\{ e^{i k_F r} f \left(\sqrt{r/\xi_1} ,
\varphi_r \xi_0 k_F/4 \right)\right\}, } \end{equation}
with $f(a,x) = \int d\phi \exp (-i\phi^2 - a | \phi -a x |)$ 
and $\xi_0 = v_F/\Delta_0$. The distance enters Eq.\ (2) as the 
ratio $r/\xi_1$ with the ``nodal'' coherence length 
$\xi_1=\frac{1}{8} k_F \xi_{0}^{2}$. This scale and its energy 
counterpart $\omega_0 = v_F/\xi_1$ arose previously in different 
problems related to the $d-$wave superconductivity \cite{AY}. 
The above integral can be reduced to the form
$$  f(a,x) =
      e^{-ia^2x^2-i\frac\pi4}\left[
      F(ae^{-i\frac\pi4} (1/2 +ix))
      +(x\to -x)\right],$$
where $F(Z)=e^{Z^2} {\rm Erfc}(Z)$. At $|Z|= a\sqrt{x^2+1/4}\ll1$ 
one has $f(a,x) = 2\sqrt{\pi}e^{-i\pi/4}$ which corresponds to 
the absence of superconducting screening near the nodes at small
distances. If $|Z|\gg 1$, the value of $F(Z)$ depends on the sign 
of Re$Z$. One can represent $f(a,x)$ for this case in the form 
$f=f^{(1)}+f^{(2)}$, where 
     \begin{eqnarray}
     f^{(1)}(a,x) &=& \frac{e^{-ia^2x^2}}{a(1+4x^2)}
     \nonumber \\
     f^{(2)}(a,x) &=&
     \sqrt{\pi} e^{-ia^2-i\frac\pi4- a^2|x|}
     \theta\left(|x|-1/2\right)
     \nonumber
     \end{eqnarray}

Balatsky and Salkola determined the angular range of unscreened 
impurity-impurity potential by matching the above expressions 
$f^{(1)}(a,x=0)$ and $f^{(2)}(a,x)$. In fact, these are two 
additive contributions from generally different domains of 
integration in (2), $\phi\simeq ax$ and $\phi\simeq 0$, 
respectively. The condition $\phi\simeq ax$ is equivalent to the 
one $\varphi\simeq\varphi_r$, which justifies our usage of
expansions for all $|\varphi_r| \lesssim 1$. We see that the
exponential term $f^{(2)}$ may exceed $f^{(1)}$ only
at $a\ll 1 \ll a|x|$ (i.e.\ at relatively small distances and
far away from node) while at $a \gtrsim 1$ the latter term   
clearly dominates. This is also confirmed by the numerical calculations. 

Hence,
 apart from the screened contribution to $\widetilde V_{ij}$ 
useless for the localization problem, at large distances $r \gtrsim \xi_1$
we have the unscreened power-law one  
{\it for all directions} of $\varphi_r$. 
This contribution can be written in the form
   
   \begin{eqnarray}
   \widetilde V_{ij}^{(1)} &\propto&  
   \frac{\sin(k_Fr|\cos\varphi_r|)}{k_Fr|\cos\varphi_r|} 
   \frac{{\Delta_0}/{\varepsilon_F}}{
   (\Delta_0/\varepsilon_F)^2\!+\!\tan^2 \varphi_r} 
   \nonumber \\ 
   && +   (\varphi_r\!\to\!\varphi_r\!+\!\pi/2),
   \nonumber
   \end{eqnarray}
where we extrapolate the angular dependences and utilize the
$\pi/2$-periodicity of Eq.\ (1). Therefore, we conclude, that
the picture of the impurity-impurity potential acting only 
within the slowly broadening tails claimed by Balatsky and Salkola 
as well as all the physical consequences of this picture appear to 
be incorrect; however, we still remain under the governing of 
Anderson's theorem \cite{A} about the absence of localization. The 
situation is changed, if we would take into account the pair-breaking 
role of impurities. For the estimates, one can incorporate this 
phenomenon adding the damping $\gamma^2$-term into the denominator 
of function $\widehat G^{0}({\bf k})$. As a result, 
$\widetilde V^{(1)}_{ij}$ also acquires the exponential 
anisotropic factor of the form $e^{-r/l(\varphi_r)}$.  
For instance, along the nodes $l(\varphi_r)\simeq l$,
where $l=v_F/\gamma$ is the mean free path of the electron.    

We thank M.\ Lavagna, P.A.\ Lee and S.V.\ Maleyev for useful 
discussions. A.Y.\ acknowledges the financial support from the 
Minist{\`e}re de l'Education Nationale de l'Enseignement 
Sup{\'e}rieur et de la Recherche, France.

\vspace{0.3cm}

 \noindent
{
D.N.\ Aristov $^{1,2}$ and A.G.\ Yashenkin $^{3,1}$
}
\vspace{0.3cm}

\noindent
{$^1$ Petersburg Nuclear Physics Institute,
Gatchina, St. Petersburg 188350, Russia.
}

\noindent
{
$^2$ Laboratoire L{\'e}on Brillouin, CE-Saclay, 
91191 Gif-sur-Yvette Cedex, France. 
}

\noindent
{
$^3$ Centre~ d'Etudes~ Nucl{\'e}aires~ de~ Grenoble, 
DRFMC/SPSMS, 17 rue des Martyrs, 38054 Grenoble Cedex 9, 
France.
}




\vspace*{-0.5cm} 
\begin{references}

\bibitem{BS} A.V.\ Balatsky and M.I.\ Salkola,
Phys.\ Rev.\ Lett.\ {\bf 76}, 2386 (1996).

\bibitem{AY}
M.L.\ Titov, A.G.\ Yashenkin, and D.N.\ Aristov,
Phys.\ Rev.\ {\bf B 52}, 10626 (1995);
A.G.\ Yashenkin, D.N.\ Aristov, and S.V.\ Maleyev,
Physica C {\bf 261}, 137, (1996);
  D.N. Aristov, S.V.  Maleyev, A.G. Yashenkin,
to appear in Z.Phys. {\bf B 102}, (1997). 


\bibitem{A} P.W.\ Anderson, Phys.\ Rev.\ {\bf 109},
1492 (1958).

\end{references}
\end{document}